\documentclass[aps,prd,12pt,superscriptaddress,nofootinbib,floatfix,longbibliography]{revtex4-2}

\usepackage{graphicx}
\usepackage{bm}
\usepackage{mathtools}
\usepackage{amsmath,amssymb}
\usepackage{orcidlink}

\hypersetup{%
  colorlinks=true,
  linkcolor=blue,
  citecolor=blue,
  urlcolor=blue,
  bookmarks=true,
  bookmarksnumbered=true,
  unicode=true}
  
\DeclareMathOperator{\arctanh}{arctanh}
\begin{document}

\title{Self-Forces as Nonlocal Probes of Gravastar Interiors}

\author{Bokai Zhang \orcidlink{0009-0004-8123-1894}}
\email{bkzhang07@163.com}
\affiliation{School of Physics, Huazhong University of Science and Technology, 1037 LuoYu Rd, Wuhan, Hubei 430074, China}

\author{Yungui Gong \orcidlink{0000-0001-5065-2259}}
\email{gongyungui@nbu.edu.cn}
\affiliation{Institute of Fundamental Physics and Quantum Technology, Department of Physics, School of Physical Science and Technology,\\
Ningbo University, 818 Fenghua Rd, Ningbo, Zhejiang 315211, China}
\affiliation{School of Physics, Huazhong University of Science and Technology, 1037 LuoYu Rd, Wuhan, Hubei 430074, China}

\date{\today}

\begin{abstract}

Compact objects with the same exterior metric are locally indistinguishable to test particles, 
yet observables built from retarded fields can retain information about the spacetime outside the particle's immediate neighbourhood. 
The self-force acting on a particle in curved spacetime provides a
unique probe of both the local geometry and the global structure of the
background spacetime.
We calculate the static, conservative self-force on minimally coupled scalar and electric charges in the simplest thin-shell gravastar: a de Sitter core matched to a Schwarzschild exterior. 
Weak-field expansions in the compactness $M/R$ are obtained analytically and
summed in closed form. For a scalar charge outside the
gravastar the self-force is nonzero---in contrast to the exactly vanishing
result for a Schwarzschild black hole of the same mass---and behaves as
$\tfrac{2}{5}q^2 M R^2/r_0^5$ at large distances, while for an electric
charge the universal Smith--Will force $e^2M/r_0^3$ is corrected by a
structure-dependent term $\tfrac{4}{5}e^2 M R^2/r_0^5$. Inside the gravastar
the scalar self-force is directed toward the center at leading order in
$M/R$, vanishes linearly at the center, and produces harmonic oscillations
of the charge about the center; the electromagnetic self-force inside the
gravastar behaves similarly. 
The results demonstrate explicitly that the self-force depends
not only on the local curvature surrounding the particle but also on the
global structure of spacetime: although the exterior geometry of a gravastar
is identical to that of a Schwarzschild black hole, the interior boundary
conditions modify the regular field and therefore produce distinct
self-forces.

\end{abstract}

\maketitle

\section{Introduction}
\label{sec:intro}

The observational identification of a compact object as a black hole ultimately requires more than measuring a Schwarzschild or Kerr exterior. Horizonless compact objects can reproduce the same exterior geometry while differing in their boundary conditions, internal stress tensor, and near-surface response. Self-force offers a complementary probe because it is sensitive not only to the local curvature at the test particle's position but also to the global structure of spacetime.
The electromagnetic self-force acting on a point charge in flat spacetime was first calculated by Dirac in 1938 \cite{Dirac:1938nz}, and was extended to curved spacetime in 1960 by DeWitt and Brehme \cite{DeWitt:1960fc,Hobbs1968}. 
The gravitational self-force was formulated independently by Mino, Sasaki, and Tanaka, and by Quinn and Wald in 1997 \cite{Mino:1996nk,Quinn:1996am}. 
Curved spacetime introduces a fundamentally new feature absent in flat spacetime. 
Because spacetime curvature scatters wave propagation, the retarded Green function develops support inside the null cone. 
The field generated by a particle therefore propagates not only along null geodesics but also through the interior of the light cone. 
The particle subsequently interacts with this scattered field, producing a nonlocal contribution known as the tail term. 
Unlike ordinary radiation reaction in flat spacetime, the self-force in curved spacetime depends on the entire past history of the particle through the geometry of the spacetime.
For reviews, see Refs. \cite{Barack:2009ux,Poisson:2011nh,Harte:2014wya,Pound:2015tma,Barack:2018yvs,LISAConsortiumWaveformWorkingGroup:2023arg}.

A point source acts on itself through the regular part of its retarded field.
In the Detweiler--Whiting description, the singular field has the same local singularity structure as the retarded field but exerts no force; the remaining regular field contains the curvature-scattered, boundary-sensitive information that changes the motion \cite{Detweiler:2002mi}. 
The resulting force is therefore not determined solely by the metric and curvature at the particle. Two spacetimes can agree throughout a neighbourhood of the worldline and nevertheless yield different conservative self-forces because their Green functions obey different global boundary conditions \cite{Burko:2000yx,Drivas:2010wz}.
A striking illustration is provided by the
spherical shell: a charge placed in the locally flat interior of the shell
experiences a nonzero self-force, while an identical charge in globally
Minkowski spacetime experiences none; 
likewise, the self-force outside the
shell differs from the one outside a black hole of the same mass, although
the exterior geometries are isometric by Birkhoff's theorem
\cite{Burko:2000yx}. 
In this sense the self-force transcends the domain of
validity of the Einstein equivalence principle, and it can be used as a probe
of the internal composition of the central body
\cite{Burko:2000yx,Isoyama:2012in,RubindeCelis:2013rai,Kumar:2019pjp}. 
At large distances the
leading piece of the self-force is universal---it depends only on the mass of
the body---while structure-dependent corrections enter at higher order in the
multipole expansion \cite{Isoyama:2012in,Seenivasan:2025ysy}.

This sensitivity to global structure makes the self-force a natural
diagnostic tool for exotic compact objects (ECOs), horizonless alternatives
to black holes whose exterior geometry can be arbitrarily close to that of a
black hole \cite{Cardoso:2019rvt}. Among the most studied ECO models is the
gravastar (gravitational vacuum condensate star) of Mazur and Mottola
\cite{Mazur:2001fv,Mazur:2004fk}, in which the would-be horizon is replaced by a
de Sitter core surrounded by a shell of ultra-relativistic matter. Visser and
Wiltshire simplified the original five-layer construction to a three-layer
model in which a de Sitter interior is joined to a Schwarzschild exterior
across a single infinitely thin shell, and showed that configurations stable
against radial perturbations exist \cite{Visser:2003ge}; subsequent work
established that gravastars require anisotropic pressures \cite{Cattoen:2005he}
and exhibited dynamically stable ``bounded excursion'' models in which the
shell oscillates between two finite radii \cite{Rocha:2008yd}. Gravastars and
other ECOs can in principle be distinguished from black holes through their
perturbation spectra and the echoes imprinted on ringdown and inspiral
waveforms \cite{Cardoso:2019rvt,Chirenti:2007mk}, a prospect that is especially
relevant for future extreme mass ratio inspiral (EMRI) observations \cite{LISAConsortiumWaveformWorkingGroup:2023arg}.

Static configurations isolate the nonlocal conservative sector without the additional complications of radiation reaction. 
An agent holding a charge at a fixed position
can therefore, in principle, read off the internal composition of the
central object from the magnitude of the force required to keep the charge
in place: a striking, if idealized, illustration of the nonlocal character
of self-interaction in curved spacetime \cite{Burko:2000yx,Isoyama:2012in}.
Static problems are the simplest setting in which this mechanism is fully
analytically tractable, and they provide benchmarks for the more involved
time-dependent computations relevant to EMRIs \cite{Bini:2008iwy,Unruh:1976fc,Burko:1999zy,Kuchar:2013bla,Leaute:1984sx,Krtous:2019uqr,Lohiya:1982fp,Leaute:1982sm,Burko:2001kr,Beach:2014aba,Taylor:2015waa,Linet:1986yp,Boisseau:1996yg,Grats:2016leu,Khusnutdinov:2010py,Krasnikov:2008kr,Khusnutdinov:2007wq,Taylor:2012mv,Bezerra:2009wj,RubindeCelis:2012fq,Khusnutdinov:2001je,BezerradeMello:1998km,Muniz:2013iua,Santos:2019dxg,Linet:1999sf,BezerradeMello:2006hz,Barbosa:2008fk,Barbosa:2010sx,BezerradeMello:2012nq,Linet:1979sj,Leaute:1985st,Vilenkin:1979im,Carvalho:2008zza,Zayats:2016iun,Mecca:2026akk}.
In this paper we compute the scalar and electromagnetic self-forces acting on
point charges held at rest either inside or outside a 
gravastar. To our knowledge this is the first self-force calculation for a
gravastar spacetime. The field equations admit exact mode solutions in terms
of hypergeometric functions in the de Sitter interior and Legendre functions
in the Schwarzschild exterior, which we match across the shell and across the
point source. Because the field of a point source diverges at the position of
the particle, the bare force must be regularized; we adopt the mode-sum
regularization prescription of Barack and Ori \cite{Barack:1999wf}, built on the
Detweiler--Whiting decomposition of the field into singular and regular parts
\cite{Detweiler:2002mi}, with the regularization parameters for static,
spherically symmetric spacetimes computed by Casals, Poisson, and Vega
\cite{Casals:2012qq}. We obtain analytic weak-field expansions in the
compactness $M/R$, sum them in closed form, and evaluate the exact
regularized mode sums numerically for arbitrary compactness.

Our main results can be summarized as follows. (i) For a scalar charge
outside the gravastar, the self-force is nonzero, in contrast to the exactly
vanishing force in the Schwarzschild black-hole spacetime; at large distances
it is repulsive and falls off as $\tfrac{2}{5}q^2 M R^2/r_0^5$, a pure
structure effect. (ii) For an electric charge outside the gravastar, the
self-force reproduces the universal Smith--Will force at leading order and
acquires a structure-dependent correction $\tfrac{4}{5}e^2 M R^2/r_0^5$,
distinct from the correction $\tfrac{2}{3}e^2 M R^2/r_0^5$ found for a hollow
shell \cite{Burko:2000yx}. (iii) Inside the gravastar the scalar self-force is
nonzero already at first order in $M/R$---in contrast to the hollow shell,
where the scalar force in the flat interior is a second post-Newtonian effect
\cite{Burko:2000yx}---is directed toward the center, and produces harmonic
oscillations about the center with frequency $\omega^2 = 2 q^2 M/(m R^4)$ to
leading order; the electromagnetic force inside the gravastar is likewise of
first order, with $\omega^2 = e^2 M/(m R^4)$, three times the value found for
a hollow shell \cite{Burko:2000yx}. (iv) In every configuration the self-force
grows without bound as the charge approaches the shell, in analogy with the
logarithmic divergence found for thin shells \cite{Burko:2000yx} and stellar
surfaces \cite{Seenivasan:2025ysy}; the divergence is an artifact of the
zero-thickness idealization. (v) In the black-hole limit, in which the
interior boundary condition is replaced by regularity at the horizon, our
formulas reproduce the known results---Wiseman's vanishing scalar force
\cite{Wiseman:2000rm} and the Smith--Will force \cite{Smith:1980tv}---and all
weak-field expansions are validated against the numerical mode sums at the
percent level or better.

The paper is organized as follows. In Sec.~\ref{sec:gravastar} we describe
the gravastar spacetime, the junction conditions at the
shell, and our setup;
review the Detweiler--Whiting decomposition and the
mode-sum regularization prescription for static spherical spacetimes.
The scalar self-force is computed in Sec.~\ref{sec:scalar}, for charges both
inside and outside the gravastar, together with the weak-field expansions,
the black-hole limit, and numerical results. Section \ref{sec:em} presents
the corresponding analysis for an electric charge, including a new
closed-form weak-field result for the interior case. In
Sec.~\ref{sec:discussion} we discuss the results,
emphasizing the role of the self-force as a probe of the global structure of
spacetime, 
and we conclude in Sec.~\ref{sec:conclusion}. 
Appendix \ref{app:identities} collects useful properties of the special
functions used in the text, and Appendix \ref{app:weakfield} presents the
weak-field expansions of the matching coefficients in detail.

Throughout we use geometrized units $G=c=1$ and the metric signature
$(-+++)$. We denote the Gauss hypergeometric function by
$F(a,b;c;z)\equiv {}_2F_1(a,b;c;z)$ and its derivative with respect to the
last argument by $F^{(1)}(a,b;c;z)$. Legendre functions of the first and
second kind are denoted $P_l$ and $Q_l$, and $f^{(n)}(a)\equiv d^nf(x)/dx^n |_{x=a}$.
We also introduce the dimensionless radial variables
\begin{equation}
x \equiv \frac{r}{M}-1, \quad x_0 \equiv \frac{r_0}{M}-1, \quad
X \equiv \frac{R}{M}-1, \quad y \equiv \frac{r_0}{R}.
\label{eq:notation}
\end{equation}

\section{The gravastar spacetime and mode-sum method}
\label{sec:gravastar}

\subsection{The Visser--Wiltshire model}\label{sec:vwmodel}

The gravastar picture was introduced by Mazur and Mottola
\cite{Mazur:2001fv,Mazur:2004fk} as a cold, endpoint configuration of gravitational
collapse that avoids both the event horizon and the central singularity of a black hole. 
In their original construction the object consists of five
layers: a de Sitter core with $p=-\rho$, a thin shell of stiff matter with
$p=\rho$, and the Schwarzschild vacuum exterior, with intervening transition
layers of continuous equation of state. Quantum-hydrodynamic analogies
motivate the de Sitter core as a vacuum condensate, and the whole structure
has no horizon: the redshift of light emitted from the surface is large but
finite. For most physical questions the transition layers are unimportant,
and Visser and Wiltshire showed that the essential features---including
dynamical stability against radial perturbations---are captured by a
three-layer model in which the de Sitter core is joined directly to the
Schwarzschild exterior across a single infinitely thin shell
\cite{Visser:2003ge}. This is the model adopted here.
The spacetime is static and spherically symmetric, with line element
\begin{equation}
ds^2 = -f(r)\,dt^2 + f(r)^{-1}\,dr^2 + r^2\,(d\theta^2+\sin^2\theta\,d\phi^2),
\label{eq:metric}
\end{equation}
where
\begin{equation}
f(r)=
\begin{cases}
1-H^2 r^2, & r<R,\\[4pt]
1-\dfrac{2M}{r}, & r>R.
\end{cases}
\label{eq:fmetric}
\end{equation}
The interior $r<R$ is a patch of de Sitter spacetime, with $H^{-1}$
the de Sitter radius, sourced by a vacuum energy with equation of state
$p=-\rho$; the exterior $r>R$ is a patch of the Schwarzschild geometry of mass
$M$. The two regions are joined at $r=R$ by an infinitely thin shell of stiff
matter, whose surface energy density and surface tension are fixed by the
Israel junction conditions \cite{Visser:2003ge,Rocha:2008yd}. Continuity of the
metric across the shell requires
\begin{equation}
1-H^2R^2 = 1-\frac{2M}{R}
\qquad\Longleftrightarrow\qquad
H^2 = \frac{2M}{R^3},
\label{eq:Hrelation}
\end{equation}
so that $H^2 R^2 = 2M/R$ is a direct measure of the compactness. The model
has no horizon provided $R>2M$, and the Buchdahl-type bound for the interior
is automatically satisfied since the de Sitter geometry is everywhere regular.
The stability of such configurations against radial perturbations was
established in Ref.~\cite{Visser:2003ge}, and dynamically stable oscillating
(``bounded excursion'') variants were constructed in Ref.~\cite{Rocha:2008yd}.
Here the shell is taken to be static at $r=R$.

On this fixed background we place a test particle of mass $m$ carrying a
scalar charge $q$ or an electric charge $e$. The particle is held at rest at
$(r_0,\theta_0,\phi_0)$ by an external force; exploiting spherical symmetry
we set $\theta_0=0$ without loss of generality. The charge is treated to
leading (test-field) order, so the self-force is $O(q^2)$ or $O(e^2)$ and the
background geometry is unaffected. We distinguish two configurations:
$r_0<R$ (charge inside the gravastar) and $r_0>R$ (charge outside the
gravastar).

\subsection{Junction conditions and the thin shell}\label{sec:junction}

The junction at $r=R$ is described by the Darmois--Israel formalism. The
induced metric on the shell must be continuous, which is guaranteed by the
matching condition \eqref{eq:Hrelation}, while the jump of the extrinsic
curvature $K_{ab}$ across the shell is related to the surface stress-energy
tensor
\begin{equation}
S_{ab} = \sigma\, u_a u_b + \vartheta\,(h_{ab}+u_a u_b)
\label{eq:Sab}
\end{equation}
by the Lanczos equation
\begin{equation}
-[K_{ab}] + [K]\,h_{ab} = 8\pi S_{ab}.
\label{eq:lanczos}
\end{equation}
Here $u^a$ is the four-velocity of the shell, $h_{ab}$ the induced metric,
$\sigma$ the surface energy density, $\vartheta$ the surface tension, and
$[X]\equiv X|_{R^+}-X|_{R^-}$ denotes the jump across the shell. Although the
metric function $f$ is continuous at $r=R$, its derivative $f'=df/dr$ is not:
$f'(R^+)=2M/R^2$ while $f'(R^-)=-2H^2R=-4M/R^2$, so that
\begin{equation}
[f'(R)] = \frac{6M}{R^2},
\label{eq:fjump}
\end{equation}
and the shell carries a nonzero surface stress-energy. Explicit expressions
for $\sigma$ and $\vartheta$ in terms of $M$, $R$, and $H$ can be found in
Refs.~\cite{Visser:2003ge,Rocha:2008yd}; for the present analysis only the
continuity of $f$ and the jump \eqref{eq:fjump} are needed. Two features of
the thin-shell idealization will be important below. First, because $f$ is
continuous, both the scalar modes $\varphi_l$ and their radial derivatives,
as well as the electromagnetic modes $a_l$ and $da_l/dr$, are continuous
across the shell: the matching conditions at $r=R$ take the simplest possible
form. Second, the discontinuity \eqref{eq:fjump} in the geometry is
ultimately responsible for the divergence of the self-force as the charge
approaches the shell, much as the surface of a constant-density star produces
a logarithmic divergence \cite{Seenivasan:2025ysy}; a shell of finite thickness
would smooth out the junction and keep the force finite.


\subsection{The Detweiler--Whiting decomposition}\label{sec:dw}

The physical self-force is most transparently defined through the
Detweiler--Whiting decomposition of the retarded field of the particle into a
singular (S) and a regular (R) part \cite{Detweiler:2002mi,Poisson:2011nh},
\begin{equation}
\Phi^{\mathrm{ret}} = \Phi^{\mathrm{S}} + \Phi^{\mathrm{R}}.
\label{eq:SRsplit}
\end{equation}
The singular field $\Phi^{\mathrm{S}}$ has a purely local construction: it
depends only on the geometry in the immediate vicinity of the particle, it
shares the singularity structure of the retarded field at the particle's
position, and it is symmetric around the particle so that it exerts no net
force on it. The regular field
$\Phi^{\mathrm{R}}=\Phi^{\mathrm{ret}}-\Phi^{\mathrm{S}}$ is a smooth
solution of the homogeneous field equation and is solely responsible for the
self-force,
\begin{equation}
f_\mu = q\,\nabla_\mu \varphi^{\mathrm{R}},
\qquad
f_\mu = e\,F^{\mathrm{R}}_{\mu\nu}u^\nu,
\label{eq:Rforce}
\end{equation}
for scalar and electric charges, respectively. Because $\Phi^{\mathrm{S}}$ is
locally constructed, identical particles at the same radius in two spacetimes
that are locally isometric possess the same singular field; any difference in
their self-forces must therefore come from $\Phi^{\mathrm{R}}$, which encodes
the global structure of spacetime through the boundary conditions.

\subsection{Mode-sum prescription}\label{sec:modesum-prescription}

In practice the retarded field is computed mode by mode. For a static
particle in a spherically symmetric spacetime, only the radial component of
the self-force is nonzero, and the mode-sum prescription of Barack and Ori
\cite{Barack:1999wf} gives
\begin{equation}
f_r = \sum_{l=0}^{\infty}\bigl(f^{\mathrm{bare}}_l - f^{\mathrm{s}}_l\bigr),
\label{eq:modesum}
\end{equation}
where $f^{\mathrm{bare}}_l$ is the $l$-mode of the bare force, computed from
the full retarded field as the average of the two one-sided radial
derivatives at the particle, and the singular modes admit the large-$l$
expansion
\begin{equation}
f^{\mathrm{s}}_l = q^2\Bigl[\tilde{A}\Bigl(l+\tfrac12\Bigr) + \tilde{B}
+ \frac{\tilde{C}}{l+\tfrac12}
+ \frac{\tilde{D}}{(l-\tfrac12)(l+\tfrac32)} + O(l^{-3})\Bigr].
\label{eq:singularmodes}
\end{equation}
The regularization parameters $\tilde{A}$, $\tilde{B}$, $\tilde{C}$, and
$\tilde{D}$ depend only on the local geometry at $r_0$ and have been computed
in closed form for arbitrary static, spherically symmetric spacetimes by
Casals, Poisson, and Vega \cite{Casals:2012qq}. The series in
Eq.~\eqref{eq:modesum} converges; subtracting the leading terms of
Eq.~\eqref{eq:singularmodes} accelerates the convergence. In each of the
cases treated below we shall display the explicit singular modes obtained
from the large-$l$ behavior of the bare modes, which agree with the general
expressions of Ref.~\cite{Casals:2012qq}.

\section{Scalar self-force}\label{sec:scalar}

\subsection{Field equation and mode decomposition}\label{sec:scalar-eq}

We consider a massless, minimally coupled scalar field $\varphi$ sourced by
the particle,
\begin{equation}
\Box\varphi = -4\pi\rho,
\qquad
\Box \coloneqq g^{\mu\nu}\nabla_\mu\nabla_\nu,
\label{eq:scalar-eom}
\end{equation}
with the charge density of a point scalar charge $q$ moving on the worldline
$z^\mu(\tau)$,
\begin{equation}
\rho = q\int_{-\infty}^{\infty}d\tau\,
\frac{\delta^{4}\bigl(x^{\mu}-z^{\mu}(\tau)\bigr)}{\sqrt{-g}}.
\label{eq:scalar-source}
\end{equation}
For a static particle at $(r_0,\theta_0=0,\phi_0=0)$ the field is static and
can be decomposed into spherical-harmonic modes,
\begin{equation}
\varphi(r,\theta,\phi) = \sum_{l,m}\varphi_{lm}(r)\,Y_{lm}(\theta,\phi).
\label{eq:scalar-modes}
\end{equation}
Since $Y_{lm}(0,\phi)\propto\delta_{m0}$, only the $m=0$ modes contribute; we
write $\varphi_l(r)\equiv\varphi_{l0}(r)$. The source mode is
\begin{equation}
\rho_l = q \sqrt{\frac{2l+1}{4\pi} f(r_0)}\,
\frac{\delta(r-r_0)}{r^2},
\label{eq:scalar-source-mode}
\end{equation}
and the radial equation following from Eq.~\eqref{eq:scalar-eom} in the
metric \eqref{eq:metric} is
\begin{equation}
\frac{1}{r^2}\frac{d}{dr}\Bigl(r^2 f\,\frac{d\varphi_l}{dr}\Bigr)
-\frac{l(l+1)}{r^2}\varphi_l
= -q\sqrt{4\pi(2l{+}1)f}\,\frac{\delta(r{-}r_0)}{r^2}.
\label{eq:scalar-radial}
\end{equation}
The mode $\varphi_l$ must be continuous everywhere, finite at $r=0$ and at
$r\to\infty$; integrating Eq.~\eqref{eq:scalar-radial} across $r=r_0$ yields
the jump condition for $d\varphi_l/dr$ at the source, while continuity of
$f$ at the shell implies that both $\varphi_l$ and $d\varphi_l/dr$ are
continuous across $r=R$.

The homogeneous equation associated with Eq.~\eqref{eq:scalar-radial} admits
exact solutions in both regions. In the de Sitter interior, the change of
variable $z=H^2r^2$ together with the ansatz $\varphi_l=(Hr)^{\sigma}u(z)$
reduces the homogeneous radial equation to the Gauss hypergeometric equation
\begin{equation}
z(1-z)u''+\Bigl[c-(a+b+1)z\Bigr]u'-ab\,u=0,
\label{eq:hypergeom-eq}
\end{equation}
with the two admissible exponents $\sigma=l$ and $\sigma=-(l+1)$, for which
$(a,b,c)=\bigl(\tfrac{l}{2},\tfrac{l+3}{2};\tfrac{2l+3}{2}\bigr)$ and
$(a,b,c)=\bigl(-\tfrac{l+1}{2},\tfrac{2-l}{2};\tfrac{1-2l}{2}\bigr)$,
respectively. Two independent solutions are therefore
\begin{equation}
(Hr)^l\,F_{s+}(H^2r^2),
\qquad
(Hr)^{-(l+1)}\,F_{s-}(H^2r^2),
\label{eq:scalar-dS-sols}
\end{equation}
where we have introduced the shorthands
\begin{align}
F_{s+}(z) &\equiv F\Bigl(\tfrac{l}{2},\tfrac{l+3}{2};\tfrac{2l+3}{2};z\Bigr),
\label{eq:Fsplus}\\
F_{s-}(z) &\equiv F\Bigl(-\tfrac{l+1}{2},\tfrac{2-l}{2};\tfrac{1-2l}{2};z\Bigr).
\label{eq:Fsminus}
\end{align}
The first solution is regular at the origin, the second is singular there.
In the Schwarzschild exterior, the change of variable $x=r/M-1$ reduces the
homogeneous radial equation to the Legendre equation
\begin{equation}
\frac{d}{dx}\Bigl[(x^2-1)\frac{d\varphi_l}{dx}\Bigr]-l(l+1)\varphi_l=0,
\label{eq:legendre-eq}
\end{equation}
whose two independent solutions are the Legendre functions
\begin{equation}
P_l\Bigl(\frac{r}{M}-1\Bigr),
\qquad
Q_l\Bigl(\frac{r}{M}-1\Bigr),
\label{eq:scalar-Schw-sols}
\end{equation}
of which $P_l$ is regular at the horizon radius $r=2M$ and $Q_l$ decays as
$r^{-(l+1)}$ at infinity. 

\subsection{Charge inside the gravastar}\label{sec:scalar-in}

Let the scalar charge be located at $r_0<R$. The mode solution of
Eq.~\eqref{eq:scalar-radial} satisfying the regularity and matching
conditions is
\begin{equation}
\varphi_l(r)=
\begin{cases}
A_l\,(Hr)^l F_{s+}(H^2r^2), & 0\leq r\leq r_0,\\[4pt]
B_l\,(Hr)^l F_{s+}(H^2r^2) & \\
\quad +\,C_l\,(Hr)^{-(l+1)} F_{s-}(H^2r^2), & r_0\leq r\leq R,\\[4pt]
D_l\,Q_l\bigl(\frac{r}{M}-1\bigr), & r\geq R.
\end{cases}
\label{eq:scalar-in-sol}
\end{equation}
Imposing continuity of $\varphi_l$ and of its derivative at $r=R$, the jump
condition at $r=r_0$, and regularity at the origin and at infinity, the four
coefficients are determined to be
\begin{align}
A_l &= \frac{C_l}{(Hr_0)^{2l+1}}
\Bigl[\Bigl(\frac{r_0}{R}\Bigr)^{2l+1}\!E_l
+\frac{F_{s-}(H^2r_0^2)}{F_{s+}(H^2r_0^2)}\Bigr],
\label{eq:scalar-in-A}\\[4pt]
B_l &= \frac{E_l\,C_l}{(HR)^{2l+1}},
\label{eq:scalar-in-B}\\[4pt]
C_l &= \frac{q}{r_0}\,(Hr_0)^{l+1}
\sqrt{\frac{4\pi(1-H^2r_0^2)}{2l+1}}\,F_{s+}(H^2r_0^2),
\label{eq:scalar-in-C}\\[4pt]
D_l &= \frac{C_l}{(HR)^{l+1}Q_l(X)}
\Bigl[E_l F_{s+}(H^2R^2)+F_{s-}(H^2R^2)\Bigr],
\label{eq:scalar-in-D}
\end{align}
where the dimensionless constant $E_l$ encodes the junction at the shell,
\begin{equation}
E_l = \frac{\bigl(l+1+\frac{R}{M}\mathcal{Q}_l\bigr)F_{s-}(H^2R^2)
-2H^2R^2 F_{s-}^{(1)}(H^2R^2)}
{\bigl(l-\frac{R}{M}\mathcal{Q}_l\bigr)F_{s+}(H^2R^2)
+2H^2R^2 F_{s+}^{(1)}(H^2R^2)},
\label{eq:scalar-in-E}
\end{equation}
with $\mathcal{Q}_l\equiv Q_l^{(1)}(X)/Q_l(X)$ the logarithmic derivative of
the Legendre function at the shell. In deriving these expressions we used the
Wronskian relations collected in Appendix \ref{app:identities}.

The only nonzero component of the self-force is the radial one. Differentiating
the mode solution \eqref{eq:scalar-in-sol} on both sides of the source and
averaging, the $l$-mode of the bare force is
\begin{align}
\bar f^{\mathrm{bare}}_l ={}&
\frac{q}{2r_0}\sqrt{\frac{2l+1}{4\pi}}
\Bigl\{(A_l+B_l)\,(Hr_0)^l
\Bigl[l F_{s+}(H^2r_0^2)
\nonumber\\
&\quad
+2H^2r_0^2 F_{s+}^{(1)}(H^2r_0^2)\Bigr]
+C_l\,(Hr_0)^{-(l+1)}
\nonumber\\
&\times
\Bigl[-(l+1) F_{s-}(H^2r_0^2)
+2H^2r_0^2 F_{s-}^{(1)}(H^2r_0^2)\Bigr]\Bigr\}.
\label{eq:scalar-in-bare}
\end{align}
Note the factor of $l$ in the first bracket, which originates from the radial
derivative of $(Hr)^l$; it guarantees that the $l=0$ mode contributes only
through the $C_l$ term and that the force vanishes at the center, as required
by spherical symmetry. At large $l$ the averaged bare modes approach the
$l$-independent constant
\begin{equation}
f^{\mathrm{s}}_l = -\frac{q^2}{2r_0^2}\,
\frac{1-2H^2r_0^2}{1-H^2r_0^2},
\label{eq:scalar-in-sing}
\end{equation}
in agreement with the general regularization parameters of
Ref.~\cite{Casals:2012qq}. The regularized
self-force is then the convergent sum
\begin{equation}
f_r = \sum_{l=0}^{\infty}
\bigl(\bar f^{\mathrm{bare}}_l - f^{\mathrm{s}}_l\bigr),
\label{eq:scalar-in-reg}
\end{equation}
in which the structure-sensitive information is carried by the matching
coefficient $E_l$ through $A_l$ and $B_l$.

\subsubsection{Weak-field expansion}\label{sec:scalar-in-weak}

An instructive analytical limit is the weak-field (small-compactness) regime
$M/R\ll1$, which implies $H^2r^2\leq H^2R^2=2M/R\ll1$. Expanding the
hypergeometric functions about $z=0$ and the Legendre functions about
$X\to\infty$ [Eqs.~\eqref{eq:hypsmallz} and \eqref{eq:QlargeX}], the
coefficient \eqref{eq:scalar-in-E} becomes
\begin{equation}
E_l = \frac{3(l+1)}{(1-2l)(1+2l)}\,\frac{M}{R}
+O\Bigl(\frac{M^2}{R^2}\Bigr),
\label{eq:scalar-in-Eweak}
\end{equation}
as derived in detail in Appendix \ref{app:weakfield}. Note that $E_l$ does
not vanish for $l=0$; the decoupling of the monopole from the force comes
instead from the explicit factor of $l$ in the derivative bracket of
Eq.~\eqref{eq:scalar-in-bare}. The regularized force \eqref{eq:scalar-in-reg}
then reduces, to leading order in $M/R$, to
\begin{equation}
f_r = \frac{q^2}{r_0^2}\,\frac{M}{R}
\sum_{l=0}^{\infty}\frac{3l(l+1)}{(1-2l)(1+2l)}\,y^{2l+1}
+O\Bigl(\frac{M^2}{R^2}\Bigr),
\label{eq:scalar-in-weak}
\end{equation}
with $y=r_0/R$. The series can be summed in closed form,
\begin{equation}
f_r = \frac{3}{8}\frac{q^2}{r_0^2}\frac{M}{R}
\Bigl[\frac{y-3y^3}{1-y^2}
-(1+3y^2)\arctanh y\Bigr]
+O\Bigl(\frac{M^2}{R^2}\Bigr).
\label{eq:scalar-in-closed}
\end{equation}
Several properties follow directly. First, the force is nonzero at first
order in $M/R$: the curvature of the de Sitter interior contributes to the
self-force already at the first post-Newtonian level. This stands in contrast
to a hollow massive shell, whose interior is flat and where the scalar
self-force is quadratic in $M/R$ \cite{Burko:2000yx}. Second, the bracket in
Eq.~\eqref{eq:scalar-in-closed} is negative for $0<y<1$: the force is
directed toward the center (the charge is repelled by the shell). Third, for
$y\to1^-$ the force diverges logarithmically, $|f_r|\sim\frac{3}{2}
\frac{q^2}{r_0^2}\frac{M}{R}|\ln(1-y)|$; we return to this surface divergence
in Sec.~\ref{sec:discussion}. Finally, near the center,
\begin{equation}
f_r = -\,2\,q^2\,\frac{M}{R^4}\,r_0
+O\Bigl(\frac{M^2}{R^2},r_0^3\Bigr),
\label{eq:scalar-in-center}
\end{equation}
so the charge undergoes harmonic oscillations about the center with angular
frequency
\begin{equation}
\omega^2 = \frac{2q^2 M}{m R^4}
+O\Bigl(\frac{M^2}{R^2}\Bigr),
\label{eq:scalar-in-omega}
\end{equation}
where $m$ is the particle's mass.

\begin{figure}[t]
\centering
\includegraphics[width=0.6\columnwidth]{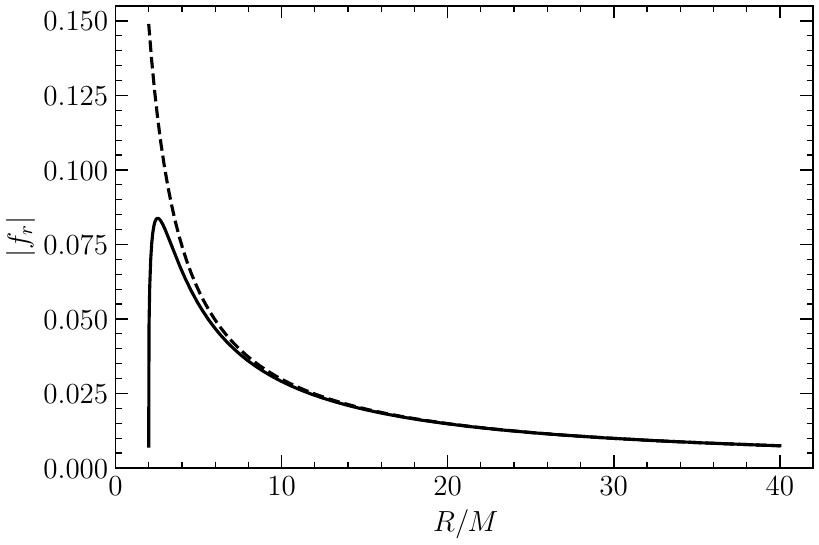}
\caption{Magnitude of the regularized radial self-force on a scalar charge
held at $r_0=R/2$ inside the gravastar, in units of $q^2/r^2_0$, as a function of the compactness
parameter $R/M$. The solid curve is obtained by numerical evaluation of the
exact mode sum \eqref{eq:scalar-in-reg}; the dashed curve is the
leading-order weak-field result \eqref{eq:scalar-in-closed}. The two agree
for $R/M\gtrsim10$ and the exact force reaches a maximum near $R/M\simeq2.5$.}
\label{fig:fsin}
\end{figure}

\subsubsection{Numerical Implementation}
The regularized sums are evaluated numerically. Two practical remarks are in
order. First, the hypergeometric and Legendre functions entering the
coefficients are computed directly from their defining series and
differential equations, with the Wronskian relations of Appendix
\ref{app:identities} used as consistency checks. Second, the sums are
truncated at a finite $l_{\text{max}}$, chosen adaptively so that the
partial sums are stable to the accuracy, we choose $l_{\text{max}}=100$. 
The exact regularized force \eqref{eq:scalar-in-reg} is plotted in
Fig.~\ref{fig:fsin} for a charge at $r_0=R/2$, together with the weak-field
expression \eqref{eq:scalar-in-closed}. The weak-field expansion accurately
reproduces the exact force for $R/M\gtrsim 10$, while for compact
configurations the exact result is significantly smaller than the weak-field
estimate and attains a maximum at $R/M\simeq 2.5$.

\subsection{Charge outside the gravastar}\label{sec:scalar-out}

Now let the scalar charge be located at $r_0>R$. The mode solution is
\begin{equation}
\varphi_l(r)=
\begin{cases}
A_l\,(Hr)^l F_{s+}(H^2r^2), & 0\leq r\leq R,\\[4pt]
B_l\,P_l\bigl(\frac{r}{M}-1\bigr)
+C_l\,Q_l\bigl(\frac{r}{M}-1\bigr), & R\leq r\leq r_0,\\[4pt]
D_l\,Q_l\bigl(\frac{r}{M}-1\bigr), & r\geq r_0,
\end{cases}
\label{eq:scalar-out-sol}
\end{equation}
with coefficients
\begin{align}
A_l &= \frac{P_l(X)+E_l Q_l(X)}
{(HR)^l F_{s+}(H^2R^2)}\,B_l,
\label{eq:scalar-out-A}\\[4pt]
B_l &= \frac{q}{M}\sqrt{4\pi(2l+1)\Bigl(1-\frac{2M}{r_0}\Bigr)}\,Q_l(x_0),
\label{eq:scalar-out-B}\\[4pt]
C_l &= E_l\,B_l,
\label{eq:scalar-out-C}\\[4pt]
D_l &= \Bigl(E_l+\frac{P_l(x_0)}{Q_l(x_0)}\Bigr)B_l,
\label{eq:scalar-out-D}
\end{align}
where
\begin{equation}
E_l = \frac{\bigl(l+2H^2R^2\mathcal{F}_l\bigr)P_l(X)
-\frac{R}{M}P_l^{(1)}(X)}
{\frac{R}{M}Q_l^{(1)}(X)
-\bigl(l+2H^2R^2\mathcal{F}_l\bigr)Q_l(X)}
\label{eq:scalar-out-E}
\end{equation}
and $\mathcal{F}_l\equiv F_{s+}^{(1)}(H^2R^2)/F_{s+}(H^2R^2)$. The
coefficient $E_l$ measures the admixture of the horizon-singular solution
$Q_l$ in the region $R<r<r_0$, induced by the interior boundary condition; it
vanishes when the interior is replaced by a black-hole horizon, since the
field regular at the horizon is proportional to $P_l$ alone.

The $l$-mode of the bare force, again evaluated as the average of the two
one-sided limits, is
\begin{align}
\bar f^{\mathrm{bare}}_l ={}&
\frac{q^2}{M^2}\sqrt{1-\frac{2M}{r_0}}\,(2l+1)
\Bigl\{\frac12\Bigl[P_l^{(1)}(x_0)Q_l(x_0)
\nonumber\\
&\quad +P_l(x_0)Q_l^{(1)}(x_0)\Bigr]
+E_l\,Q_l(x_0)Q_l^{(1)}(x_0)\Bigr\}.
\label{eq:scalar-out-bare}
\end{align}
After subtracting the singular modes \eqref{eq:singularmodes} with the
parameters of Ref.~\cite{Casals:2012qq} for the Schwarzschild exterior, the part
of the mode sum that does not involve $E_l$ vanishes identically---this is
the content of Wiseman's result that the self-force on a static scalar charge
in the Schwarzschild black-hole spacetime is exactly zero
\cite{Wiseman:2000rm}. The entire regularized self-force is therefore carried by
the structure-sensitive coefficient $E_l$,
\begin{equation}
f_r = \frac{q^2}{M^2}\sqrt{1-\frac{2M}{r_0}}
\sum_{l=0}^{\infty}(2l+1)\,E_l\,Q_l(x_0)\,Q_l^{(1)}(x_0).
\label{eq:scalar-out-reg}
\end{equation}

\subsubsection{Weak-field expansion}\label{sec:scalar-out-weak}

Expanding Eq.~\eqref{eq:scalar-out-reg} in powers of $M/R$ (see Appendix
\ref{app:weakfield} for the expansion of the matching coefficients), the
leading term is
\begin{equation}
f_r = \frac{q^2}{r_0^2}\frac{M}{R}
\sum_{l=0}^{\infty}\frac{3l(l+1)}{(1+2l)(3+2l)}\,y^{-(2l+1)}
+O\Bigl(\frac{M^2}{R^2}\Bigr),
\label{eq:scalar-out-weak}
\end{equation}
which can be summed to the closed form
\begin{equation}
f_r = \frac{q^2}{r_0^2}\frac{M}{R}
\left[\frac{9y^3-3y}{8(y^2-1)}
-\frac{3}{8}(3y^2+1)\arctanh y^{-1}\right]
+O\left(\frac{M^2}{R^2}\right).
\label{eq:scalar-out-closed}
\end{equation}
The $l=0$ mode does not contribute, in accord with the absence of a universal
leading term for scalar charges \cite{Isoyama:2012in,Seenivasan:2025ysy}. The force
is positive (repulsive, directed away from the gravastar) and diverges
logarithmically as $y\to1^+$. At large distances,
\begin{equation}
f_r = \frac{2}{5}\,\frac{q^2 M R^2}{r_0^5}
+O\Bigl(\frac{R^4}{r_0^7},\frac{M^2}{R^2}\Bigr),
\label{eq:scalar-out-far}
\end{equation}
a pure structure effect: it vanishes with the size of the object and would be
absent altogether for a black hole. Note also the different parametric
dependence compared with the hollow shell, for which the exterior scalar
force is $\frac13 q^2 M^2 R/r_0^5$, a second post-Newtonian effect
\cite{Burko:2000yx}; the de Sitter interior produces a force of first order in
$M/R$.

\begin{figure}[t]
\centering
\includegraphics[width=0.6\columnwidth]{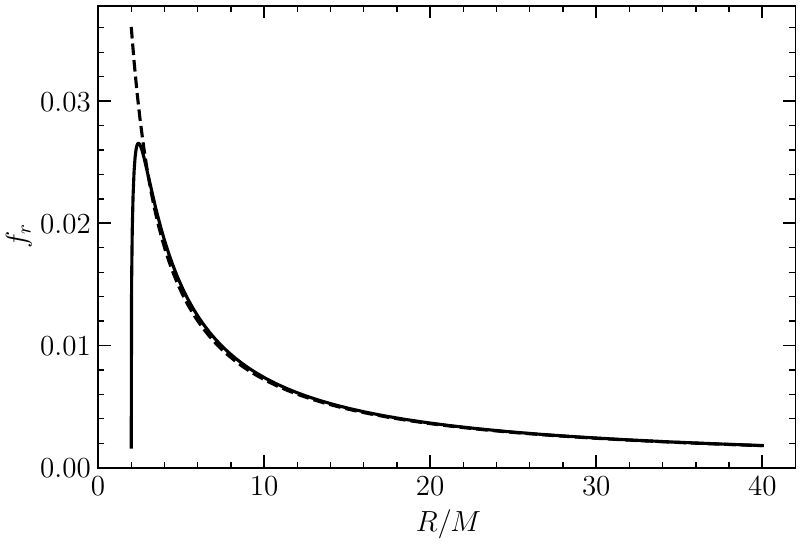}
\caption{Regularized radial self-force on a scalar charge held at $r_0=2R$
outside the gravastar, in units of $q^2/r^2_0$, as a function of the compactness parameter $R/M$. The
solid curve is the exact mode sum \eqref{eq:scalar-out-reg}; the dashed curve
is the leading-order weak-field result \eqref{eq:scalar-out-closed}. The
force is everywhere repulsive.}
\label{fig:fsout}
\end{figure}

Figure~\ref{fig:fsout} shows the exact mode sum \eqref{eq:scalar-out-reg} for
a charge at $r_0=2R$ together with the weak-field expression
\eqref{eq:scalar-out-closed}. The force is repulsive for all compactnesses,
and the weak-field expansion provides a good approximation for
$R/M\gtrsim 10$.

\subsubsection{Two special limits}\label{sec:scalar-limits}

Two limiting positions of the charge deserve a separate comment. The first
is the center, $r_0\to0$. Spherical symmetry requires the force to vanish
there, and the weak-field result \eqref{eq:scalar-in-center} shows that it
does so linearly, $f_r\simeq -2q^2Mr_0/R^4$: the center is a stable
equilibrium point for the scalar charge, with the oscillation frequency
\eqref{eq:scalar-in-omega}. Mode by mode, only $l=0$ contributes at the
center, and the $l=0$ term of Eq.~\eqref{eq:scalar-in-bare} vanishes with
the derivative of the regular solution. The second limit is the shell,
$r_0\to R^\pm$, where the force diverges logarithmically on both sides
[Eqs.~\eqref{eq:scalar-in-closed} and \eqref{eq:scalar-out-closed}]. The
divergence is not a failure of the regularization---the mode sums converge
for every $r_0\neq R$---but a genuine property of the thin-shell geometry:
the jump \eqref{eq:fjump} in $f'$ makes the local geometry at the shell
singular, and the field of a charge placed arbitrarily close to it is
arbitrarily distorted. A realistic gravastar with a shell of finite thickness
$\delta\ll R$ would regularize the divergence at $|r_0-R|\sim\delta$
\cite{Burko:2000yx,Seenivasan:2025ysy}.

A third, more subtle limit is the ultracompact one, $R\to2M^+$. For an
exterior observer the gravastar then becomes operationally indistinguishable
from a black hole: the redshift of the shell diverges and the light-crossing
time of the interior cavity grows logarithmically. Whether the self-force
approaches the black-hole result in this limit is not obvious, because the
interior boundary condition remains that of the de Sitter core rather than
horizon absorption; the matching coefficient $E_l$ does not have an
elementary limit. We leave the detailed analysis of the ultracompact limit
to future work.

\subsubsection{The black-hole limit}\label{sec:scalar-out-bh}

A valuable consistency check is obtained by replacing the de Sitter interior
with a black-hole horizon. Regularity of the static field at the horizon
$r=2M$ selects the solution proportional to $P_l(r/M-1)$ alone in the region
$2M<r<r_0$, since $Q_l$ diverges logarithmically at $x=1$; equivalently, the
black-hole boundary condition amounts to setting $E_l=0$ in
Eqs.~\eqref{eq:scalar-out-sol}--\eqref{eq:scalar-out-E}. The regularized
force \eqref{eq:scalar-out-reg} then vanishes identically: we have verified
numerically that the mode sum of the $E_l$-independent terms in
Eq.~\eqref{eq:scalar-out-bare}, after subtraction of its $l$-independent
large-$l$ limit, vanishes to within the accuracy of the computation. 
This is precisely Wiseman's result that a static,
minimally coupled scalar charge in the Schwarzschild black-hole spacetime
experiences no self-force \cite{Wiseman:2000rm}. The gravastar result
\eqref{eq:scalar-out-reg} can therefore be read directly as the
\emph{departure} from the black-hole force: the entire effect is the
boundary-condition term. It is instructive that the sign of the force is
repulsive; informally, the de Sitter core ``reflects'' part of the field back
toward the charge, whereas the horizon absorbs it.

\section{Electromagnetic self-force}\label{sec:em}

\subsection{Field equation and mode decomposition}\label{sec:em-eq}

We now consider an electric charge $e$ held at rest at $r_0$. The field
strength $F_{\mu\nu}=\partial_\mu A_\nu-\partial_\nu A_\mu$ satisfies the
Maxwell equation
\begin{equation}
\nabla_\nu F^{\mu\nu} = 4\pi j^\mu,
\label{eq:maxwell}
\end{equation}
with the current of a point charge on the worldline $z^\alpha(\tau)$,
\begin{equation}
j^\mu = e\int_{-\infty}^{\infty}d\tau\,u^\mu\,
\frac{\delta^4\bigl(x^\alpha-z^\alpha(\tau)\bigr)}{\sqrt{-g}}.
\label{eq:em-source}
\end{equation}
For a static charge only $u^t$ is nonzero, and one may consistently take the
only nontrivial component of the potential to be $A_t$. The current mode is
\begin{equation}
j^t = \frac{e}{r^2}\,\delta(r-r_0)
\sum_{l,m}Y_{lm}(\theta,\phi)\,Y_{lm}^{*}(\theta_0,\phi_0),
\label{eq:em-jt}
\end{equation}
and decomposing
\begin{equation}
A_t(r,\theta,\phi)=\sum_{l,m}a_{lm}(r)\,Y_{lm}(\theta,\phi)
\label{eq:em-modes}
\end{equation}
(with $a_l\equiv a_{l0}$ the only contributing modes for $\theta_0=0$), the
$\mu=t$ component of Eq.~\eqref{eq:maxwell} reduces to
\begin{equation}
\frac{1}{r^2}\frac{d}{dr}\Bigl(r^2\frac{da_l}{dr}\Bigr)
-\frac{l(l+1)}{f\,r^2}\,a_l
= \frac{e}{r^2}\sqrt{4\pi(2l+1)}\,\delta(r-r_0).
\label{eq:em-radial}
\end{equation}
The potential $a_l$ is continuous everywhere and finite at $r=0$ and at
infinity; integrating Eq.~\eqref{eq:em-radial} across the shell shows that
$da_l/dr$ is continuous at $r=R$, while across the source one finds
\begin{equation}
a_l^{(1)}(r_0^{+})-a_l^{(1)}(r_0^{-})
=\frac{e}{r_0^2}\sqrt{4\pi(2l+1)}.
\label{eq:em-jump}
\end{equation}

The homogeneous equation associated with Eq.~\eqref{eq:em-radial} has exact
solutions
\begin{equation}
(Hr)^l F_{v+}(H^2r^2),
\qquad
(Hr)^{-(l+1)} F_{v-}(H^2r^2)
\label{eq:em-dS-sols}
\end{equation}
in the de Sitter interior, obtained as in Sec.~\ref{sec:scalar-eq} through
the change of variable $z=H^2r^2$, with
\begin{align}
F_{v+}(z) &\equiv F\Bigl(\tfrac{l+1}{2},\tfrac{l}{2};\tfrac{2l+3}{2};z\Bigr),
\label{eq:Fvplus}\\
F_{v-}(z) &\equiv F\Bigl(-\tfrac{l+1}{2},-\tfrac{l}{2};\tfrac{1-2l}{2};z\Bigr),
\label{eq:Fvminus}
\end{align}
and, in the Schwarzschild exterior,
\begin{equation}
(r-2M)\,P_l^{(1)}\Bigl(\frac{r}{M}-1\Bigr),
\qquad
(r-2M)\,Q_l^{(1)}\Bigl(\frac{r}{M}-1\Bigr).
\label{eq:em-Schw-sols}
\end{equation}
The exterior solutions follow from the change of variable $x=r/M-1$: writing
$a_l=(x-1)w_l(x)$, the homogeneous radial equation reduces to the associated
Legendre equation of order one for $w_l$, whose independent solutions
$P_l^{(1)}$ and $Q_l^{(1)}$ give Eq.~\eqref{eq:em-Schw-sols}. Note that the
exterior electromagnetic modes involve the derivatives of the Legendre
functions, in contrast to the scalar case \eqref{eq:scalar-Schw-sols}; this
difference ultimately traces back to the vector character of the field and
is responsible for the different large-distance behavior of the two forces.

\subsection{Charge inside the gravastar}\label{sec:em-in}

For $r_0<R$ the mode solution is
\begin{equation}
a_l(r)=
\begin{cases}
A_l\,(Hr)^l F_{v+}(H^2r^2), & 0\leq r\leq r_0,\\[4pt]
B_l\,(Hr)^l F_{v+}(H^2r^2) & \\
\quad +\,C_l\,(Hr)^{-(l+1)} F_{v-}(H^2r^2), & r_0\leq r\leq R,\\[4pt]
D_l\,(r-2M)\,Q_l^{(1)}\bigl(\frac{r}{M}-1\bigr), & r\geq R,
\end{cases}
\label{eq:em-in-sol}
\end{equation}
with
\begin{align}
A_l &= \Bigl[\frac{E_l}{(HR)^{2l+1}}
+\frac{F_{v-}(H^2r_0^2)}
{(Hr_0)^{2l+1}F_{v+}(H^2r_0^2)}\Bigr]C_l,
\label{eq:em-in-A}\\[4pt]
B_l &= \frac{E_l\,C_l}{(HR)^{2l+1}},
\label{eq:em-in-B}\\[4pt]
C_l &= -\frac{e}{r_0}\,(Hr_0)^{l+1}
\sqrt{\frac{4\pi}{2l+1}}\,F_{v+}(H^2r_0^2),
\label{eq:em-in-C}\\[4pt]
D_l &= \frac{E_l F_{v+}(H^2R^2)+F_{v-}(H^2R^2)}
{(HR)^{l+1}(R-2M)\,Q_l^{(1)}(X)}\,C_l,
\label{eq:em-in-D}
\end{align}
and
\begin{widetext}
\begin{equation}
E_l = \frac{\Bigl(1+l+\dfrac{1}{1-2M/R}+\dfrac{R}{M}\widetilde{\mathcal{Q}}_l\Bigr)
F_{v-}(H^2R^2)-2H^2R^2 F_{v-}^{(1)}(H^2R^2)}
{\Bigl(l-\dfrac{1}{1-2M/R}-\dfrac{R}{M}\widetilde{\mathcal{Q}}_l\Bigr)
F_{v+}(H^2R^2)+2H^2R^2 F_{v+}^{(1)}(H^2R^2)},
\label{eq:em-in-E}
\end{equation}
\end{widetext}
where $\widetilde{\mathcal{Q}}_l\equiv Q_l^{(2)}(X)/Q_l^{(1)}(X)$ is the
logarithmic derivative of $Q_l^{(1)}$ at the shell. The Wronskian identity
\eqref{eq:wronskian-hyp} was used to simplify these expressions.

The $l$-mode of the bare radial force, evaluated as the average of the two
one-sided limits at $r=r_0$, is
\begin{align}
f^{\mathrm{bare}}_l ={}&
\frac{e^2}{r_0^2}\frac{2l+1}{2}\frac{1}{\sqrt{1-H^2r_0^2}}
-\frac{e^2}{r_0^2}\frac{1}{\sqrt{1-H^2r_0^2}}
\nonumber\\
&\times\Bigl[\Bigl(\frac{r_0}{R}\Bigr)^{2l+1}E_l F_{v+}(H^2r_0^2)
+F_{v-}(H^2r_0^2)\Bigr]
\nonumber\\
&\times\Bigl[2H^2r_0^2 F_{v+}^{(1)}(H^2r_0^2)+l F_{v+}(H^2r_0^2)\Bigr].
\label{eq:em-in-bare}
\end{align}
The regularized force follows from the mode sum \eqref{eq:modesum} upon
subtracting the singular modes. The averaged bare modes
\eqref{eq:em-in-bare} approach the $l$-independent constant
\begin{equation}
f^{\mathrm{s}}_l = \frac{e^2}{2r_0^2}\,\frac{1}{1-H^2r_0^2}
\label{eq:em-in-sing}
\end{equation}
at large $l$, again in agreement with the general parameters of
Ref.~\cite{Casals:2012qq}, and the remaining series
$f_r=\sum_l(\bar f^{\mathrm{bare}}_l-f^{\mathrm{s}}_l)$ converges and is
evaluated numerically.

\subsubsection{Weak-field expansion}\label{sec:em-in-weak}

As in the scalar case, the weak-field regime $M/R\ll1$ admits a closed-form
treatment. Expanding the coefficient \eqref{eq:em-in-E} to leading order
(Appendix \ref{app:weakfield}) gives
\begin{equation}
E_l = \frac{3l}{(2l-1)(2l+1)}\,\frac{M}{R}
+O\Bigl(\frac{M^2}{R^2}\Bigr),
\label{eq:em-in-Eweak}
\end{equation}
and the regularized force reduces to
\begin{equation}
f_r = -\frac{e^2}{r_0^2}\,\frac{M}{R}
\sum_{l=0}^{\infty}\frac{3l^2}{(2l-1)(2l+1)}\,y^{2l+1}
+O\Bigl(\frac{M^2}{R^2}\Bigr),
\label{eq:em-in-weak}
\end{equation}
with the closed-form sum
\begin{equation}
f_r = -\frac{3e^2}{8r_0^2}\,\frac{M}{R}
\Bigl[\frac{2y}{1-y^2}-y-(1-y^2)\arctanh y\Bigr]
+O\Bigl(\frac{M^2}{R^2}\Bigr).
\label{eq:em-in-closed}
\end{equation}
The bracket is positive for $0<y<1$, so the force is directed toward the
center, and it diverges logarithmically as $y\to1^-$. The $l=0$ mode does not
contribute, and near the center
\begin{equation}
f_r = -\,e^2\,\frac{M}{R^4}\,r_0
+O\Bigl(\frac{M^2}{R^2},r_0^3\Bigr),
\label{eq:em-in-center}
\end{equation}
so the charge oscillates harmonically about the center with frequency
\begin{equation}
\omega^2 = \frac{e^2 M}{m R^4}
+O\Bigl(\frac{M^2}{R^2}\Bigr).
\label{eq:em-in-omega}
\end{equation}
This is exactly three times the frequency found by Burko, Liu, and Soen for
an electric charge inside a hollow massive shell, $\omega^2=e^2M/(3mR^4)$
\cite{Burko:2000yx}: the curved de Sitter interior and the flat shell interior
are already distinguishable by the leading weak-field coefficient.

\begin{figure}[t]
\centering
\includegraphics[width=0.6\columnwidth]{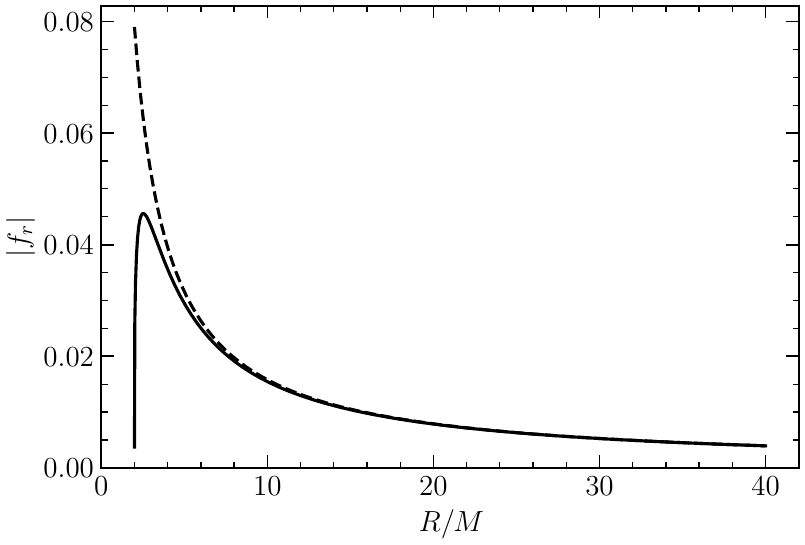}
\caption{Magnitude of the regularized radial self-force on an electric charge
held at $r_0=R/2$ inside the gravastar, in units of $e^2/r^2_0$, as a function of the compactness
parameter $R/M$ (solid curve). The dashed curve shows the weak-field
expression \eqref{eq:em-in-closed}. As in the scalar case
(Fig.~\ref{fig:fsin}), the exact force
reaches a maximum near $R/M\simeq2.5$ and decreases for more compact
configurations.}
\label{fig:fvin}
\end{figure}

Figure~\ref{fig:fvin} displays the regularized force on an electric charge at
$r_0=R/2$. The qualitative
behavior mirrors the scalar case: the force
vanishes at the center, grows toward the
shell, and, for fixed $r_0/R$, reaches a maximum at a compactness
$R/M\simeq2.5$.

\subsection{Charge outside the gravastar}\label{sec:em-out}

For $r_0>R$ the monopole mode must be treated separately. Solving
Eq.~\eqref{eq:em-radial} for $l=0$ gives
\begin{equation}
a_0(r)=
\begin{cases}
-\sqrt{4\pi}\,\dfrac{e}{r_0}, & r\leq r_0,\\[6pt]
-\sqrt{4\pi}\,\dfrac{e}{r}, & r\geq r_0,
\end{cases}
\label{eq:em-out-l0}
\end{equation}
so the monopole potential is insensitive to the presence of the shell. For
$l\geq1$ the mode solution is
\begin{equation}
a_l(r)=
\begin{cases}
A_l\,(Hr)^l F_{v+}(H^2r^2), & 0\leq r\leq R,\\[4pt]
B_l\,(r-2M)P_l^{(1)}\bigl(\frac{r}{M}-1\bigr) & \\
\quad +\,C_l\,(r-2M)Q_l^{(1)}\bigl(\frac{r}{M}-1\bigr), & R\leq r\leq r_0,\\[4pt]
D_l\,(r-2M)\,Q_l^{(1)}\bigl(\frac{r}{M}-1\bigr), & r\geq r_0,
\end{cases}
\label{eq:em-out-sol}
\end{equation}
with
\begin{align}
A_l &= \frac{(R-2M)\bigl[P_l^{(1)}(X)+E_l Q_l^{(1)}(X)\bigr]}
{(HR)^l F_{v+}(H^2R^2)}\,B_l,
\label{eq:em-out-A}\\[4pt]
B_l &= \frac{e}{M^3}\,\frac{r_0-2M}{l(l+1)}\,Q_l^{(1)}(x_0),
\label{eq:em-out-B}\\[4pt]
C_l &= E_l\,B_l,
\label{eq:em-out-C}\\[4pt]
D_l &= \Bigl(E_l+\frac{P_l^{(1)}(x_0)}{Q_l^{(1)}(x_0)}\Bigr)B_l,
\label{eq:em-out-D}
\end{align}
where
\begin{widetext}
\begin{equation}
E_l = \frac{\bigl[1-\bigl(l+2H^2R^2\widetilde{\mathcal{F}}_l\bigr)(1-2M/R)\bigr]
P_l^{(1)}(X)+(R/M-2)\,P_l^{(2)}(X)}
{\bigl[\bigl(l+2H^2R^2\widetilde{\mathcal{F}}_l\bigr)(1-2M/R)-1\bigr]
Q_l^{(1)}(X)-(R/M-2)\,Q_l^{(2)}(X)},
\label{eq:em-out-E}
\end{equation}
\end{widetext}
with $\widetilde{\mathcal{F}}_l\equiv F_{v+}^{(1)}(H^2R^2)/F_{v+}(H^2R^2)$.

The $l$-modes of the bare force are
\begin{align}
f^{\mathrm{bare}}_l ={}&
\frac{e^2}{r_0^2}\,\frac{2l+1}{l(l+1)}
\bigl(\frac{r_0}{M}\bigr)^{3}\sqrt{1-\frac{2M}{r_0}}\Bigl\{P_l^{(1)}(x_0)Q_l^{(1)}(x_0)
+E_l Q_l^{(1)}(x_0)Q_l^{(1)}(x_0)\nonumber\\
&\qquad
+\left(\frac{r_0}{M}-2\right)\left[
\frac12 P_l^{(2)}(x_0)Q_l^{(1)}(x_0)
+\frac12 P_l^{(1)}(x_0)Q_l^{(2)}(x_0)
+E_l Q_l^{(1)}(x_0)Q_l^{(2)}(x_0)\right]\Bigr\}
\label{eq:em-out-bare}
\end{align}
for $l\geq1$, and
\begin{equation}
f^{\mathrm{bare}}_0 = \frac{e^2}{2r_0^2}
\Bigl(1-\frac{2M}{r_0}\Bigr)^{-1/2}
\label{eq:em-out-bare0}
\end{equation}
for the monopole. The large-$l$ limit of the bare modes yields the singular
part
\begin{equation}
f^{\mathrm{s}}_l = \frac{e^2}{2r_0^2}\,
\frac{1-3M/r_0}{1-2M/r_0},
\label{eq:em-out-sing}
\end{equation}
which agrees with the general regularization parameters of
Ref.~\cite{Casals:2012qq} specialized to the Schwarzschild exterior. The
regularized force is $f_r=\sum_{l=0}^{\infty}(f^{\mathrm{bare}}_l-f^{\mathrm{s}}_l)$.
In this sum, the part that does not involve $E_l$ reproduces the Smith--Will
force \cite{Smith:1980tv},
\begin{equation}
f^{\mathrm{SW}}_{\hat r} = \frac{e^2 M}{r_0^3},
\label{eq:smithwill}
\end{equation}
as measured in the orthonormal frame of a static observer; the remainder,
proportional to $E_l$, encodes the internal structure of the gravastar.

\subsubsection{Weak-field expansion}\label{sec:em-out-weak}

Expanding the regularized force to leading order in $M/R$ (Appendix
\ref{app:weakfield}) gives
\begin{equation}
f_r = \frac{e^2}{r_0^2}\,\frac{M}{R}
\sum_{l=0}^{\infty}\frac{3(1+l)^2}{(1+2l)(3+2l)}\,y^{-(2l+1)}
+O\Bigl(\frac{M^2}{R^2}\Bigr),
\label{eq:em-out-weak}
\end{equation}
with the closed-form sum
\begin{equation}
f_r = \frac{e^2}{r_0^2}\,\frac{3M}{8R}
\Bigl[\frac{y^3+y}{y^2-1}
+(1-y^2)\arctanh y^{-1}\Bigr]
+O\Bigl(\frac{M^2}{R^2}\Bigr).
\label{eq:em-out-closed}
\end{equation}
Unlike the scalar case, the $l=0$ mode contributes here, reflecting the
universal character of the electromagnetic self-force at leading order. At
large distances,
\begin{equation}
f_r = \frac{e^2 M}{r_0^3}
\Bigl[1+\frac{4}{5}\Bigl(\frac{R}{r_0}\Bigr)^{2}
+O\Bigl(\frac{R^4}{r_0^4},\frac{M}{R}\Bigr)\Bigr]:
\label{eq:em-out-far}
\end{equation}
the leading term is the Smith--Will force, and the first structure-dependent
correction is $\frac45 e^2 M R^2/r_0^5$, to be compared with the correction
$\frac23 e^2 M R^2/r_0^5$ for a hollow shell of the same mass and radius
\cite{Burko:2000yx}. The expression \eqref{eq:em-out-closed} diverges as
$y\to1^{+}$: at first order in $M/R$ the self-force grows without bound as
the charge approaches the shell from outside.

\begin{figure}[t]
\centering
\includegraphics[width=0.6\columnwidth]{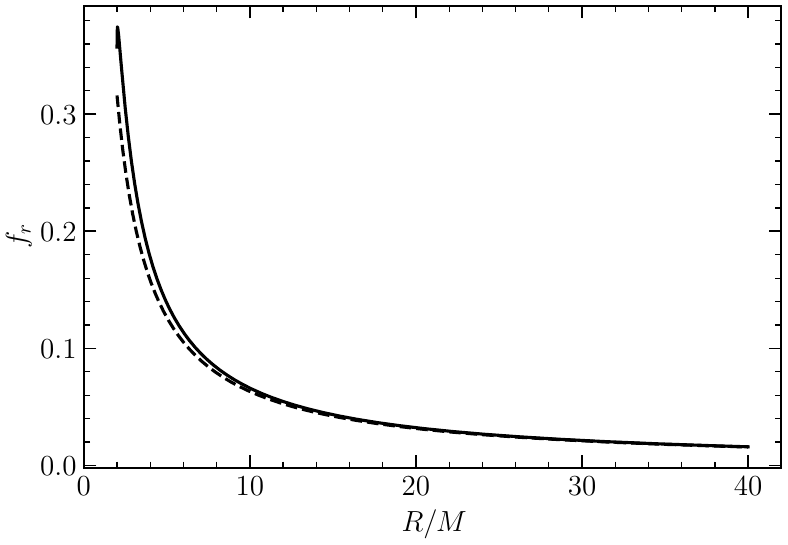}
\caption{Regularized radial self-force on an electric charge held at $r_0=2R$
outside the gravastar, in units of $e^2/r^2_0$, as a function of the compactness parameter $R/M$
(solid curve), together with the leading-order weak-field result
\eqref{eq:em-out-closed} (dashed curve). The force is everywhere repulsive
and decreases monotonically with $R/M$.}
\label{fig:fvout}
\end{figure}

Figure~\ref{fig:fvout} shows the exact mode sum for a charge at $r_0=2R$
together with the weak-field expression. The force is repulsive and
monotonically decreasing with $R/M$; the weak-field expansion captures the
exact result for $R/M\gtrsim 10$.

\subsubsection{The black-hole limit and the Copson--Linet solution}
\label{sec:em-out-bh}

As in the scalar case, setting $E_l=0$ in
Eqs.~\eqref{eq:em-out-sol}--\eqref{eq:em-out-E} amounts to imposing the
black-hole boundary condition at the horizon. 
The mode solution then
reproduces, mode by mode, the classic closed-form solution for a point charge
in the Schwarzschild spacetime, first obtained by Copson \cite{copson1928}
and corrected by Linet \cite{Linet:1976sq},
\begin{equation}
A_t(r,\theta) = \frac{e}{rr_0}\,
\frac{(r-M)(r_0-M)-M^2\cos\theta}{D(r,\theta)},
\label{eq:copsonlinet}
\end{equation}
with
\begin{equation}
\left[D(r,\theta)\right]^2\equiv\left[(r-M)(r_0-M)-M^2\cos\theta\right]^2
-M^2(r-2M)(r_0-2M)\sin^2\theta,
\label{eq:copsonlinetD}
\end{equation}
from which Smith and Will derived the self-force
$f^{\mathrm{SW}}_{\hat r}=e^2M/r_0^3$ by subtracting the locally flat
Coulomb singularity and evaluating the regular field at the charge
\cite{Smith:1980tv}. Our mode-sum computation reproduces this result when
$E_l=0$, confirming both the normalization of the mode solutions and the
regularization constants. For the gravastar, $E_l\neq0$ and the additional
term in the regularized force,
\begin{equation}
\Delta f_r \equiv f_r - f_r^{\mathrm{SW}},
\label{eq:em-out-delta}
\end{equation}
is a pure interior-structure effect. In the weak-field regime it is read off
from Eqs.~\eqref{eq:em-out-closed} and \eqref{eq:em-out-far}: at large
distances $\Delta f_r \simeq \frac45 e^2 M R^2/r_0^5$, distinct from the
$\frac23 e^2 M R^2/r_0^5$ produced by a hollow shell of the same mass and
radius \cite{Burko:2000yx}.

Compared with the scalar case,
the exterior force decays more slowly, $r_0^{-3}$ versus $r_0^{-5}$, because
of the universal Smith--Will leading term; the interior profiles are
qualitatively similar, with the electromagnetic force roughly half the scalar
one in magnitude at weak compactness [compare Eqs.~\eqref{eq:scalar-in-center}
and \eqref{eq:em-in-center}].

\section{Discussion}\label{sec:discussion}

\begin{figure}[t]
\centering
\includegraphics[width=0.6\columnwidth]{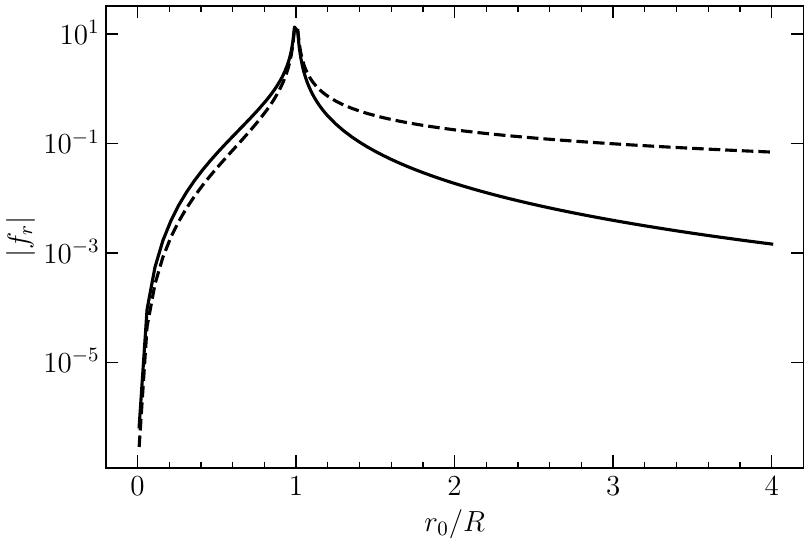}
\caption{Magnitude of the regularized radial self-force as a function of the
charge position $r_0/R$ for a gravastar with $R=4M$, in units of $q^2(e^2)/r^2_0$, on a logarithmic scale.
The solid curve is the scalar self-force, the dashed curve the
electromagnetic one. Both forces vanish linearly at the center and diverge as
the charge approaches the shell at $r_0=R$ from either side. Outside the
gravastar the electromagnetic force decays as $r_0^{-3}$ (the Smith--Will
law), while the scalar force decays as $r_0^{-5}$.}
\label{fig:ftotal}
\end{figure}

Figure~\ref{fig:ftotal} summarizes our results for a compact gravastar with
$R=4M$: it displays the regularized self-force on scalar and electric charges
across the whole domain $0<r_0/R<4$. Several features deserve comment.

\emph{Dependence on the global structure.} The exterior geometry of the
gravastar is locally isometric to that of a Schwarzschild black hole of the
same mass (Birkhoff's theorem), and the interior is locally de Sitter. Yet
the self-forces differ from their black-hole counterparts in every respect:
the scalar force outside the gravastar is nonzero [Eq.~\eqref{eq:scalar-out-reg}]
while it vanishes identically outside a Schwarzschild black hole
\cite{Wiseman:2000rm}; the electromagnetic force differs from the Smith--Will
force by structure-dependent terms; and a scalar charge inside the gravastar
feels a force of order $M/R$ even though a static worldline in the de Sitter
interior is a geodesic. As emphasized in Ref.~\cite{Burko:2000yx}, these facts
violate a naive application of the equivalence principle: the self-force is a
nonlocal effect, sensitive to boundary conditions imposed at the shell and at
infinity, and in principle it can be used to infer the internal composition
of the central object \cite{Isoyama:2012in,Kumar:2019pjp}.

\emph{Comparison with other interiors.} It is instructive to contrast the
gravastar with the hollow massive shell of Ref.~\cite{Burko:2000yx}, whose
interior is flat. For a scalar charge inside a shell the force is a second
post-Newtonian effect, $O\bigl((M/R)^2\bigr)$, whereas inside the gravastar
the curvature of the de Sitter core produces a force of order $M/R$. Outside,
the gravastar's scalar force $\frac25 q^2 M R^2/r_0^5$ is likewise of first
order in $M/R$, while the shell's is $\frac13 q^2 M^2 R/r_0^5$. For electric
charges the exterior forces share the universal Smith--Will leading term, but
the structure corrections differ: $\frac45 e^2 M R^2/r_0^5$ for the gravastar
versus $\frac23 e^2 M R^2/r_0^5$ for the shell. The self-force thus
distinguishes between interiors that are observationally degenerate at the
level of the exterior metric, in line with the general multipole analysis of
Ref.~\cite{Isoyama:2012in} and with recent results for the constant-density star
\cite{Seenivasan:2025ysy}.

\emph{Comparison with the constant-density star.} The Schwarzschild star, a
uniform-density fluid ball matched to the Schwarzschild exterior, is the
best-studied horizonless interior in this context
\cite{Shankar:2007xx,Seenivasan:2025ysy}. Shankar and Whiting computed the force on
a static electric charge outside such a star and found that it exceeds the
Smith--Will force \cite{Shankar:2007xx}; Seenivasan and Dolan revisited the
problem with both scalar and electromagnetic fields, confirmed the
enhancement, and showed that the force diverges logarithmically as the charge
approaches the stellar surface \cite{Seenivasan:2025ysy}. Our results parallel
these findings but reveal a qualitative difference in the interior physics.
In the constant-density star the matter sources an $O(\rho)$ modification of
the interior geometry already at leading order, whereas the de Sitter core of
the gravastar is an exact solution of the vacuum Einstein equations with a
cosmological constant: the interior curvature is entirely of vacuum origin.
Nevertheless the same organizing principles apply---the exterior force
decomposes into a universal part and a structure-dependent part suppressed by
powers of $R/r_0$ \cite{Isoyama:2012in}, and the interior force vanishes at the
center and diverges at the boundary. The coefficient of the logarithmic
surface divergence, as well as the magnitude of the structure-dependent
corrections, depends on the equation of state of the interior and could in
principle be used to distinguish between these competing models.

\emph{Structure coefficients.} The cleanest way to compare interiors is
through the far-field expansion of the exterior force, which we write as
$f_r=f_r^{\mathrm{univ}}+\kappa\,(\mathrm{charge})^2 MR^2/r_0^5+\cdots$,
where $f_r^{\mathrm{univ}}$ is the universal, mass-only part and $\kappa$ is
a dimensionless structure coefficient. Table~\ref{tab:structure} collects
the results for the gravastar, the hollow shell \cite{Burko:2000yx}, and the
Schwarzschild black hole. Three observations are in order. First, the scalar
universal term vanishes, so the scalar force is entirely a structure effect;
the gravastar's $\kappa_s=\frac25$ is moreover of \emph{first} post-Newtonian
order in $M/R$, while the shell's is of second order, reflecting the
curvature of the de Sitter core. Second, the electromagnetic universal term
is the Smith--Will force, and the structure correction distinguishes the de
Sitter interior ($\kappa_v=\frac45$) from the flat one ($\kappa_v=\frac23$).
Third, all structure effects scale as $MR^2/r_0^5$---the self-force analog of
a quadrupole moment \cite{Isoyama:2012in}---so they are most prominent for
extended, non-compact configurations.

\begin{table}[htp]
\caption{Far-field radial self-force outside different central objects of
mass $M$ and radius $R$, to leading order in $M/R$ and $R/r_0$. The hollow
shell results are from Ref.~\cite{Burko:2000yx}; the black-hole results are
those of Refs.~\cite{Wiseman:2000rm,Smith:1980tv}.}
\label{tab:structure}
\begin{ruledtabular}
\begin{tabular}{lcc}
Object & Scalar & Electromagnetic \\
\hline
Gravastar & $\frac{2}{5}q^2\dfrac{MR^2}{r_0^5}$ &
$\dfrac{e^2M}{r_0^3}\Bigl[1+\dfrac{4}{5}\Bigl(\dfrac{R}{r_0}\Bigr)^2\Bigr]$ \\[10pt]
Hollow shell & $\frac{1}{3}q^2\dfrac{M^2R}{r_0^5}$ &
$\dfrac{e^2M}{r_0^3}\Bigl[1+\dfrac{2}{3}\Bigl(\dfrac{R}{r_0}\Bigr)^2\Bigr]$ \\[10pt]
Black hole & $0$ & $\dfrac{e^2M}{r_0^3}$ \\
\end{tabular}
\end{ruledtabular}
\end{table}

\emph{Surface divergence.} Both forces diverge as $r_0\to R$ from either
side; the weak-field expressions
\eqref{eq:scalar-in-closed}, \eqref{eq:scalar-out-closed},
\eqref{eq:em-in-closed}, and
\eqref{eq:em-out-closed} exhibit the logarithmic character of the divergence
through the $\arctanh$ terms. The same phenomenon occurs for thin shells
\cite{Burko:2000yx} and at the surface of a Schwarzschild star
\cite{Seenivasan:2025ysy}, where it was traced to the boundary discontinuity of
the background. Physically, the divergence is an artifact of the
zero-thickness idealization: a shell of finite thickness, or a smooth
matching layer, resolves the surface and keeps the self-force finite
\cite{Burko:2000yx,Seenivasan:2025ysy}. 
The gravastar force also vanishes linearly
at the center, so that a displaced charge oscillates harmonically about
$r_0=0$ [Eqs.~\eqref{eq:scalar-in-omega} and \eqref{eq:em-in-omega}].

\emph{Oscillations about the center.} The vanishing of the interior forces
at $r_0=0$ is not merely a consistency check: it makes the center a stable
equilibrium point for both scalar and electric charges, with the harmonic
frequencies \eqref{eq:scalar-in-omega} and \eqref{eq:em-in-omega}. The
oscillation period is
\begin{equation}
T = 2\pi\sqrt{\frac{mR^4}{\kappa\,(\text{charge})^2 M}}
= 2\pi\sqrt{\frac{R^3}{M}}\,
\sqrt{\frac{mR}{\kappa\,(\text{charge})^2}},
\label{eq:oscperiod}
\end{equation}
where $\kappa=2$ and $(\text{charge})^2=q^2$ for a scalar charge, and
$\kappa=1$ and $(\text{charge})^2=e^2$ for an electric one. The first factor
in the second form is the dynamical (free-fall) timescale of the gravastar,
and the second involves the charge-to-mass ratio: unless
$(\text{charge})^2/m$ is a sizable fraction of $R$, the self-force
oscillation is slow compared with the dynamical time. In the extreme
situation $(\text{charge})^2/m\sim M$ and $R\sim10M$,
Eq.~\eqref{eq:oscperiod} gives $T\sim4\times10^2M$, comparable to the
dynamical period $2\pi\sqrt{R^3/M}\simeq2\times10^2M$. Two comments are in
order. First, the static computation captures only the conservative
restoring force; a charge released off center would in reality radiate
scalar or electromagnetic waves, and the oscillation would slowly damp as
the charge settles toward the center---a dissipative effect accessible only
in the time-dependent problem. Second, the very existence of a stable
equilibrium tied to the self-force is a structure effect: it relies on the
reflection of the field by the shell, encoded in the coefficient $E_l$, and
it has no analog for a charge near a black hole, where no interior
equilibrium exists.

\emph{Implications for gravitational self-forces and EMRIs.} Although our
computation concerns test scalar and electromagnetic fields, the mechanism it
exhibits is generic: the regular field---and hence the self-force---inherits
the boundary conditions of the spacetime. For a particle orbiting an ECO, the
tail part of the Green function, which drives the gravitational self-force,
scatters off the interior structure (or its absence of horizon) and therefore
differs from its black-hole value. This is the same physics that produces
gravitational-wave echoes in the ringdown of ECOs
\cite{Chirenti:2007mk,Cardoso:2019rvt}, and it implies that EMRI waveforms around
gravastar-like objects carry imprints of the interior at the level of the
conservative dynamics, not only of the dissipative fluxes. Static
calculations such as the present one provide analytically controlled
benchmarks for these effects.

\emph{Higher multipoles and measurability.} It is worth emphasizing the
multipole structure of our results. The universal leading term of the
exterior force ($e^2M/r_0^3$ for the electric case, zero for the scalar case)
measures only the mass of the central object, while the leading
structure-dependent correction scales as $MR^2/r_0^5$ in both cases. This
hierarchy mirrors the multipole expansion of the external field of an
extended body: the correction is the self-force analog of a quadrupole
effect, and its coefficient ($\frac45$ for the gravastar, $\frac23$ for the
shell) is a direct fingerprint of the interior \cite{Isoyama:2012in}. For an
EMRI around a putative gravastar, the analogous structure-dependent terms in
the gravitational self-force would accumulate over the $\sim10^5$ orbital
cycles in the LISA band, so that even small deviations from the black-hole
boundary condition can become observable in the phase of the waveform
\cite{Barack:2009ux,LISAConsortiumWaveformWorkingGroup:2023arg}. 

\section{Conclusions}\label{sec:conclusion}

We have computed the scalar and electromagnetic self-forces on point charges
held at rest inside or outside a thin-shell gravastar, combining exact
mode solutions in terms of hypergeometric and Legendre functions with
mode-sum regularization based on the Detweiler--Whiting singular field. The
main findings are: (i) a scalar charge outside the gravastar experiences a
nonzero repulsive self-force, $\frac25 q^2 M R^2/r_0^5$ at large distances,
in stark contrast with the exactly vanishing black-hole result; (ii) an
electric charge outside experiences the Smith--Will force corrected by the
structure-dependent term $\frac45 e^2 M R^2/r_0^5$; (iii) inside the
gravastar the forces are nonzero at first order in the compactness, directed
toward the center, and produce harmonic oscillations about it, with
frequencies $\omega^2=2q^2M/(mR^4)$ (scalar) and $\omega^2=e^2M/(mR^4)$
(electric), summarized by the closed-form weak-field expressions
\eqref{eq:scalar-in-closed} and \eqref{eq:em-in-closed}; (iv) the
forces diverge logarithmically as the charge approaches the thin shell, an
artifact of the idealized zero-thickness junction; and (v) in the black-hole
limit the construction reduces to the known results of
Refs.~\cite{Wiseman:2000rm,Smith:1980tv,copson1928,Linet:1976sq}, and the analytic
weak-field expansions agree with the numerical mode sums at the percent level
or better.
In conclusion, the conservative regular field is not fixed by the local Schwarzschild geometry
and it also depends on the inner boundary condition. 

It is worth summarizing the unifying picture that emerges. In every
configuration studied here, the difference between the gravastar and the
black hole is controlled by a single ingredient: the inner boundary
condition. Mode by mode, that condition is encoded in the reflection
coefficient $E_l$, which vanishes for horizon absorption and is nonzero for
the de Sitter core. The structure-dependent self-forces, the oscillations
about the center, the far-field structure coefficients of
Table~\ref{tab:structure}, and the logarithmic surface divergence are all,
ultimately, different facets of the same reflective cavity; in the
time-dependent problem the same cavity produces the ringdown echoes. The
static self-force is thus best viewed as a zero-frequency probe of the
cavity's reflectivity.

Natural extensions of this work include the self-force in gravastar models
with a shell of finite thickness or with anisotropic-pressure layers
\cite{Cattoen:2005he}, which would regularize the surface divergence; rotating
gravastar spacetimes; and, most importantly, the genuinely time-dependent
problem of a charge in motion. The latter would give access to the
dissipative part of the self-force and make direct contact with the physics
of echoes and of EMRI waveforms.

\begin{acknowledgments}
This research is supported in part by the National Natural Science Foundation of China under Grant Nos. 12535002 and 12588101.
\end{acknowledgments}

\appendix

\section{Special-function identities}\label{app:identities}

We collect here the properties of the hypergeometric and Legendre functions
used in the main text. The hypergeometric function is defined by 
\begin{equation}
    F(a,b;c;z) = \sum_{n=0}^{\infty} \frac{(a)_n (b)_n}{(c)_n n!}z^n,
\end{equation}
where
\begin{equation}
    (a)_0 = 1,
    \quad (a)_n = a(a+1)(a+2) \cdots (a+n-1),n \geq 1.
\end{equation}
The derivative of the Gauss hypergeometric function is
\begin{equation}
F^{(1)}(a,b;c;z)\equiv\frac{dF}{dz}
=\frac{ab}{c}\,F(a+1,b+1;c+1;z),
\label{eq:hyperiv}
\end{equation}
and the two independent solutions of the hypergeometric equation satisfy the
Wronskian relation
\begin{align}
W\left\{F(a,b;c;z),\,
z^{1-c}F(a-c+1,b-c+1;2-c;z)\right\}
=(1-c)\,z^{-c}(1-z)^{c-a-b-1}.
\label{eq:wronskian-hyp}
\end{align}
For small argument,
\begin{equation}
F(a,b;c;z) = 1+\frac{ab}{c}\,z+O(z^2).
\label{eq:hypsmallz}
\end{equation}
The Legendre functions are
\begin{equation}
P_l(x) = F\left(l+1,-l;1;\tfrac{1-x}{2}\right),
\end{equation}
\begin{equation}
Q_l(x) = \frac{\sqrt{\pi}\,\Gamma(l+1)}
{\Gamma\bigl(l+\tfrac32\bigr)\,(2x)^{l+1}}F\left(\frac{l+2}{2},\frac{l+1}{2};\frac{2l+3}{2};\frac{1}{x^2}\right),
\end{equation}
which satisfy the following relations,
\begin{equation}
W\bigl\{P_l(x),Q_l(x)\bigr\}=\frac{1}{1-x^2},
\label{eq:wronskian-leg}
\end{equation}
\begin{equation}
W\bigl\{P_l^{(1)}(x),Q_l^{(1)}(x)\bigr\}=\frac{l(l+1)}{(x^2-1)^2},
\end{equation}
and at large argument the function of the second kind behaves as
\begin{equation}
Q_l(x) = c_l\,x^{-(l+1)}
\Bigl[1+\frac{(l+1)(l+2)}{2(2l+3)}\,x^{-2}+O(x^{-4})\Bigr],
\label{eq:QlargeX}
\end{equation}
with
\begin{equation}
c_l \equiv \frac{\sqrt{\pi}\,\Gamma(l+1)}
{2^{l+1}\,\Gamma\bigl(l+\tfrac32\bigr)},
\qquad x\to\infty.
\label{eq:cldef}
\end{equation}
Using the duplication formula for the Gamma function, $c_l$ can be
rewritten in terms of the leading coefficient
$d_l\equiv(2l)!/\bigl(2^l(l!)^2\bigr)$ of $P_l(x)=d_lx^l[1+O(x^{-2})]$ as
\begin{equation}
c_l = \frac{1}{(2l+1)\,d_l}.
\label{eq:cldl}
\end{equation}

\section{Weak-field expansions}\label{app:weakfield}

In this appendix we derive the leading weak-field behavior of the matching
coefficients $E_l$ and of the regularized forces. We write
\begin{equation}
\epsilon \equiv \frac{M}{R}\ll1,
\qquad
H^2R^2=2\epsilon,
\qquad
H^2r_0^2 = 2\epsilon y^2,
\label{eq:epsdef}
\end{equation}
and expand all quantities to first order in $\epsilon$.

\subsection{Useful expansions}

For the Legendre function at the shell, $X=R/M-1$ and
$X^{-1}=\epsilon(1+\epsilon+\cdots)$, so Eq.~\eqref{eq:QlargeX} gives the
logarithmic derivatives
\begin{align}
\frac{R}{M}\,\mathcal{Q}_l
&\equiv \frac{R}{M}\frac{Q_l^{(1)}(X)}{Q_l(X)}
= -(l+1)(1+\epsilon)+O(\epsilon^2),
\label{eq:calQexp}\\
\frac{R}{M}\,\widetilde{\mathcal{Q}}_l
&\equiv \frac{R}{M}\frac{Q_l^{(2)}(X)}{Q_l^{(1)}(X)}
= -(l+2)(1+\epsilon)+O(\epsilon^2).
\label{eq:calQtexp}
\end{align}
The hypergeometric functions are expanded with Eq.~\eqref{eq:hypsmallz},
\begin{align}
F(a,b;c;z)&=1+\frac{ab}{c}z+O(z^2),
\nonumber\\
F^{(1)}(a,b;c;z)&=\frac{ab}{c}+O(z),
\label{eq:hypsmallzB}
\end{align}
with $z=H^2R^2=2\epsilon$ at the shell and $z=H^2r_0^2=O(\epsilon)$ at the
charge.

\subsection{Scalar charge inside}

For the scalar inside coefficient \eqref{eq:scalar-in-E} we need
\begin{equation}
\frac{ab}{c}\Big|_{s-}
= \frac{(l+1)(l-2)}{2(1-2l)},
\qquad
\frac{ab}{c}\Big|_{s+}
= \frac{l(l+3)}{2(2l+3)}.
\label{eq:abcs}
\end{equation}
Inserting Eqs.~\eqref{eq:calQexp} and \eqref{eq:hypsmallzB} into the
numerator of Eq.~\eqref{eq:scalar-in-E},
\begin{align}
N_l &= \bigl(l+1-(l+1)(1+\epsilon)\bigr)\bigl(1+O(\epsilon)\bigr)
-4\epsilon\,\frac{(l+1)(l-2)}{2(1-2l)}+O(\epsilon^2)
\nonumber\\
&= \epsilon\Bigl[-(l+1)-\frac{2(l+1)(l-2)}{1-2l}\Bigr]+O(\epsilon^2)
\nonumber\\
&= \frac{3(l+1)}{1-2l}\,\epsilon+O(\epsilon^2),
\label{eq:appNscalar}
\end{align}
while the denominator is
\begin{equation}
D_l = \bigl(l+(l+1)(1+\epsilon)\bigr)\bigl(1+O(\epsilon)\bigr)+O(\epsilon)
= (2l+1)+O(\epsilon),
\label{eq:appDscalar}
\end{equation}
so that
\begin{equation}
E_l = \frac{3(l+1)}{(1-2l)(2l+1)}\,\epsilon+O(\epsilon^2),
\label{eq:appEscalarin}
\end{equation}
which is Eq.~\eqref{eq:scalar-in-Eweak}. In the bare modes
\eqref{eq:scalar-in-bare}, the coefficient $E_l$ enters only through the
combination $(r_0/R)^{2l+1}E_l$ in $A_l$; all remaining pieces are
$l$-dependent but independent of the shell and cancel against the singular
constant \eqref{eq:scalar-in-sing} order by order in $\epsilon$. The
derivative bracket contributes its explicit factor of $l$, and one arrives at
Eq.~\eqref{eq:scalar-in-weak},
\begin{equation}
f_r = \frac{q^2}{r_0^2}\,\epsilon
\sum_{l=0}^{\infty}\frac{3l(l+1)}{(1-2l)(1+2l)}\,y^{2l+1}.
\label{eq:appscalarinsum}
\end{equation}
Using the partial-fraction decomposition
\begin{equation}
\frac{3l(l+1)}{(1-2l)(1+2l)}
= -\frac34+\frac{9}{8}\,\frac{1}{1-2l}
-\frac{3}{8}\,\frac{1}{1+2l},
\label{eq:appscalardecomp}
\end{equation}
the series is summed by
$\sum_{l=0}^\infty y^{2l+1}/(2l+1)=\arctanh y$ and
$\sum_{l=0}^\infty y^{2l+1}/(1-2l)=y-y^2\arctanh y$, giving the closed form
\eqref{eq:scalar-in-closed}.

\subsection{Electric charge inside}

For the electromagnetic coefficient \eqref{eq:em-in-E}, note that
$1/(1-2M/R)=1+2\epsilon+O(\epsilon^2)$ and
\begin{equation}
\frac{ab}{c}\Big|_{v-}
= \frac{l(l+1)}{2(1-2l)},
\qquad
\frac{ab}{c}\Big|_{v+}
= \frac{l(l+1)}{2(2l+3)}.
\label{eq:abcv}
\end{equation}
With Eq.~\eqref{eq:calQtexp} the numerator becomes
\begin{align}
N_l &= \bigl(1+l+1+2\epsilon-(l+2)(1+\epsilon)\bigr)\bigl(1+O(\epsilon)\bigr)
-4\epsilon\,\frac{l(l+1)}{2(1-2l)}+O(\epsilon^2)
\nonumber\\
&= \epsilon\Bigl[-l-\frac{2l(l+1)}{1-2l}\Bigr]+O(\epsilon^2)
= \frac{3l}{2l-1}\,\epsilon+O(\epsilon^2),
\label{eq:appNem}
\end{align}
and the denominator
\begin{equation}
D_l = \bigl(l-1+(l+2)\bigr)\bigl(1+O(\epsilon)\bigr)+O(\epsilon)
= (2l+1)+O(\epsilon),
\label{eq:appDem}
\end{equation}
so
\begin{equation}
E_l = \frac{3l}{(2l-1)(2l+1)}\,\epsilon+O(\epsilon^2),
\label{eq:appEemin}
\end{equation}
which is Eq.~\eqref{eq:em-in-Eweak}. The factor of $l$ in the derivative
bracket of Eq.~\eqref{eq:em-in-bare} then yields
Eq.~\eqref{eq:em-in-weak},
\begin{equation}
f_r = -\frac{e^2}{r_0^2}\,\epsilon
\sum_{l=0}^{\infty}\frac{3l^2}{(2l-1)(2l+1)}\,y^{2l+1},
\label{eq:appeminsum}
\end{equation}
and the decomposition
\begin{equation}
\frac{3l^2}{(2l-1)(2l+1)}
= \frac34+\frac{3}{8}\Bigl(\frac{1}{2l-1}-\frac{1}{2l+1}\Bigr)
\label{eq:appemdecomp}
\end{equation}
sums the series to the closed form \eqref{eq:em-in-closed}. In particular,
$3l^2/[(2l-1)(2l+1)]\sim 3/4$ at large $l$, so the sum inherits the
logarithmic divergence of $\sum y^{2l+1}$ as $y\to1^-$; and the small-$y$
behavior $f_r\simeq -e^2 M r_0/R^4$ follows from the $l=1$ term.

\subsection{Charges outside}

The outside coefficients, Eqs.~\eqref{eq:scalar-out-E} and
\eqref{eq:em-out-E}, are expanded along the same lines, using in addition
$P_l(X)=d_l X^l\bigl[1+O(X^{-2})\bigr]$ with
$d_l=(2l)!/\bigl(2^l(l!)^2\bigr)$. At leading order the numerators are
dominated by the balance between the $P_l$ and $P_l^{(1)}$ terms, which
cancels at $O(1)$ and leaves an $O(\epsilon)$ remainder, while the
denominators are dominated by $(R/M)Q_l^{(1)}(X)$. Carrying out the
expansion one finds, for $l\geq1$,
\begin{equation}
E_l = -\,\frac{3l}{2l+3}\,d_l^{\,2}
\Bigl(\frac{R}{M}\Bigr)^{2l}\bigl[1+O(\epsilon)\bigr],
\label{eq:appEout}
\end{equation}
for both the scalar and the electromagnetic coefficient: at leading
order the reflection of the higher modes off the shell is independent of
the spin of the field. We have verified this statement, including the
coefficient in Eq.~\eqref{eq:appEout}, numerically to better than twelve
significant digits at $R/M=3\times10^3$. The scalar monopole is an
exception: the numerator of Eq.~\eqref{eq:scalar-out-E} vanishes
identically for $l=0$, since $F_{s+}=1$ and $P_0^{(1)}=0$, so that
$E_0=0$ exactly; the electromagnetic monopole is likewise insensitive to
the shell [Eq.~\eqref{eq:em-out-l0}]. This explains why the exterior
forces receive their leading structure-dependent contributions from the
$l\geq1$ modes.

The growth of $E_l$ as $(R/M)^{2l}$ in Eq.~\eqref{eq:appEout} is only an
apparent obstacle to the weak-field limit: it is compensated in the force
sums by the decay of the Legendre functions at the charge. From
Eqs.~\eqref{eq:QlargeX} and \eqref{eq:cldl},
\begin{equation}
Q_l(x_0)\,Q_l^{(1)}(x_0)
= -\,\frac{l+1}{(2l+1)^2d_l^{\,2}}
\Bigl(\frac{M}{r_0}\Bigr)^{2l+3}
\biggl[1+O\Bigl(\frac{M}{r_0}\Bigr)\biggr],
\label{eq:appQQ}
\end{equation}
and inserting Eq.~\eqref{eq:appEout} together with \eqref{eq:appQQ} into
the mode sum \eqref{eq:scalar-out-reg} yields directly the coefficient
$3l(l+1)/\bigl[(1+2l)(3+2l)\bigr]$ of Eq.~\eqref{eq:scalar-out-weak}. The
same computation gives the structure-dependent part of the electromagnetic
series \eqref{eq:em-out-weak}; its remaining, universal part is the
mode-by-mode expansion of the Smith--Will force. 
The resulting series are easy to calculate and their large-distance
limits, Eqs.~\eqref{eq:scalar-out-far} and \eqref{eq:em-out-far}, follow
from $\arctanh(y^{-1})=y^{-1}+ y^{-3}/3+ y^{-5}/5+\cdots$.


%

\end{document}